Sankar Das Sarma is a physics faculty member at the University of Maryland in College Park. Dong-Ling Deng is an assistant professor and Lu-Ming Duan is a CC Yao Professor in the Institute for Interdisciplinary Information Sciences at Tsinghua University in Beijing.

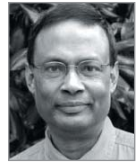 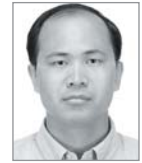 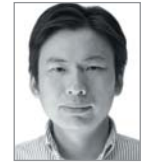

# MACHINE LEARNING *meets* QUANTUM PHYSICS

Sankar Das Sarma, Dong-Ling Deng, and Lu-Ming Duan

**The marriage of the two fields may give birth to a new research frontier that could transform them both.**

Machine learning is a field of computer science that seeks to build computers capable of discovering meaningful information and making predictions about data. It is the core of artificial intelligence (AI) and has powered many aspects of modern technologies, from face recognition and natural language processing to automated self-driving cars.

The field is rapidly growing, and its applications have become ubiquitous.[1] Google Translate's online service uses machine learning to convert Chinese characters into English text with no human intervention. Machine-learning techniques were recently used to build AlphaGo,[2] a robot that has defeated the world's best players in Go, an ancient board game; developers have considered mastering the game as the highest AI achievement. Until AlphaGo demonstrated its prowess, the game was widely thought to be too intricate for machines to excel at because of the huge number of possible moves.

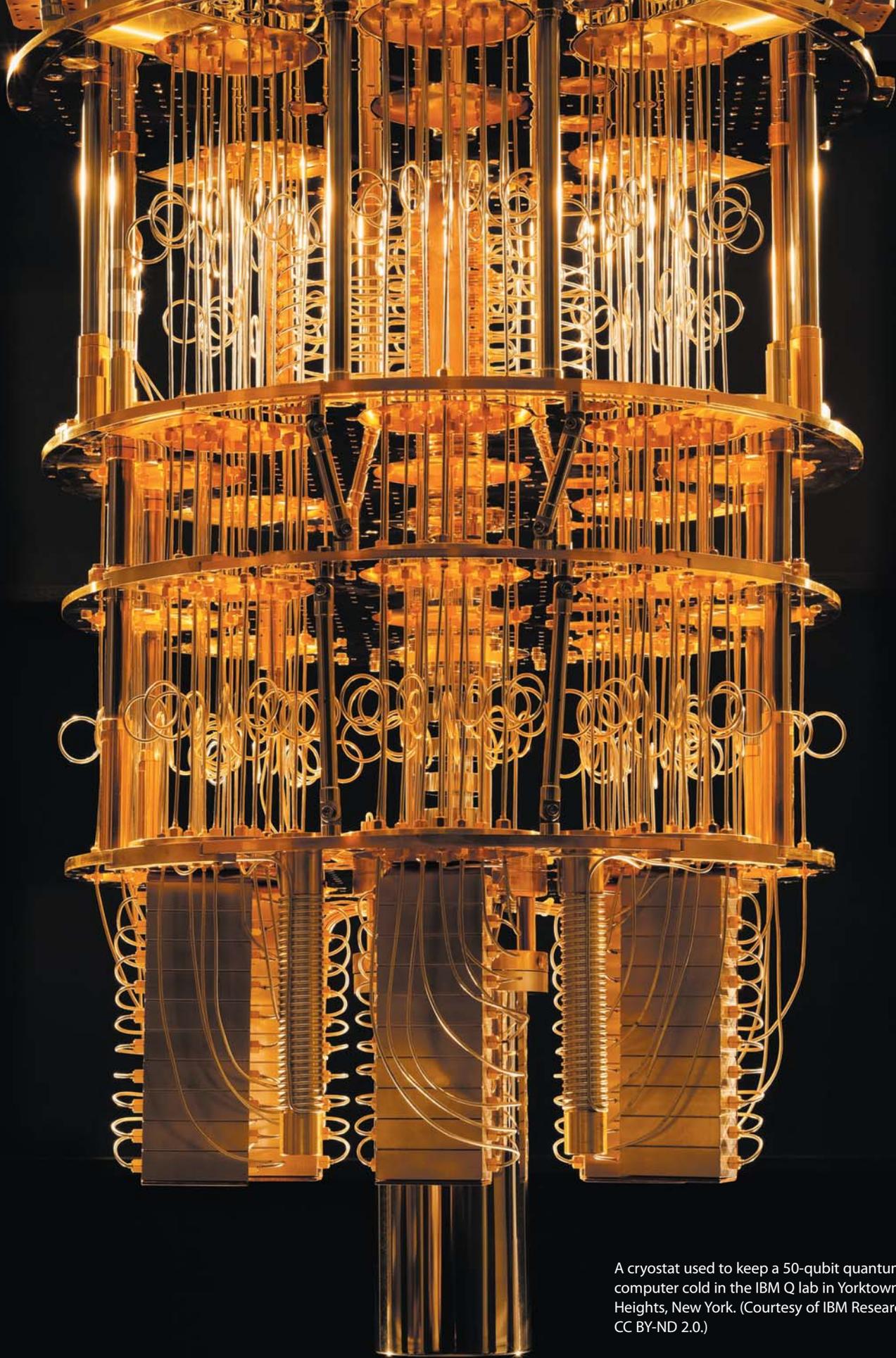

A cryostat used to keep a 50-qubit quantum computer cold in the IBM Q lab in Yorktown Heights, New York. (Courtesy of IBM Research, CC BY-ND 2.0.)

One of the biggest problems facing machine learning is the so-called curse of dimensionality—in general the number of training data sets required for the machine to learn the desired information is exponential in the dimension $d$. If a data set lies in a high-dimensional space, then it quickly becomes computationally unmanageable. That complexity is similar to quantum mechanics, for which an exponential amount of information is generally also required to fully describe a quantum many-body state.

Despite its intricacies, quantum theory is arguably the most successful quantitative theory of nature. It not only provides the basis for understanding physics on all length scales, from elementary particles like electrons and quarks to gigantic objects like stars and galaxies, but also lays the foundation for modern technologies ranging from lasers and transistors to nuclear magnetic resonators and even quantum computers.[3] Given the great successes of both machine learning and quantum physics, one may ask: Can these two seemingly unrelated but intimately connected fields merge in a seamless, synergistic manner?

It sounds like science fiction, but that fusion is happening right now and may lead to presently unimaginable breakthroughs in both fields. Machine learning has progressed dramatically over the past two decades, and many problems that were extremely challenging or even inaccessible to automated learning have now been solved. Those successes raise new possibilities for machine learning to solve open problems in quantum physics.

Meanwhile, the idea of quantum information processing has revolutionized theories and implementations of computation. New quantum algorithms may offer tantalizing prospects to enhance machine learning itself. The interaction between machine learning and quantum physics will undoubtedly benefit both fields.

## Uncovering phases of matter

When applying machine learning to physics problems, a straightforward strategy is to use supervised learning, in which an algorithm is trained with data that are labeled beforehand; the algorithm's goal is to take that information and establish a general rule for assigning labels to data outside the training set. For example, in identifying pictures of dogs and cats, a supervised learning algorithm will take thousands of images labeled either "dog" or "cat" and determine a relationship between the images' pixel values and their labels. It then assigns those labels to images that it has not seen before.

The same supervised learning technique can be used for identifying distinct phases of matter and the transitions between them, one of the central problems in condensed-matter physics. Juan Carrasquilla and Roger Melko were the first to explore that idea in their study of the ferromagnetic Ising model, which features discrete atomic spins arranged on a lattice.[4] The spins display a disordered paramagnetic phase at high temperatures and an ordered ferromagnetic phase at low temperatures, and a phase transition occurs between the two at some critical temperature $T_c$.

Instead of sorting dogs and cats, Carrasquilla and Melko used equilibrium spin configurations sampled from Monte Carlo simulations to train the algorithm to identify paramagnetic and ferromagnetic states. They demonstrated that after training with those labeled samples, the algorithm could correctly assign the labels to new samples. Moreover, by scanning a range of temperatures, it located $T_c$ and found the critical exponents that are crucial to the study of phase transitions.

Supervised learning requires that users know *a priori* how their data should be categorized. Alternatively, unsupervised learning uses unlabeled training data and allows the network to find meaningful patterns and structures in them. A common example of unsupervised learning is clustering, in which training data are divided into several groups based on identified similarities and those groups are used to categorize new, previously unseen data. In 2016 Lei Wang applied clustering to the Ising model and successfully identified the paramagnetic and ferromagnetic phases and the transition between them, despite not giving the algorithm explicit sorting criteria.[5] Around the same time, Evert van Nieuwenburg and coworkers proposed a confusion scheme that combined both supervised and unsupervised learning.[6] They tested their approach on several

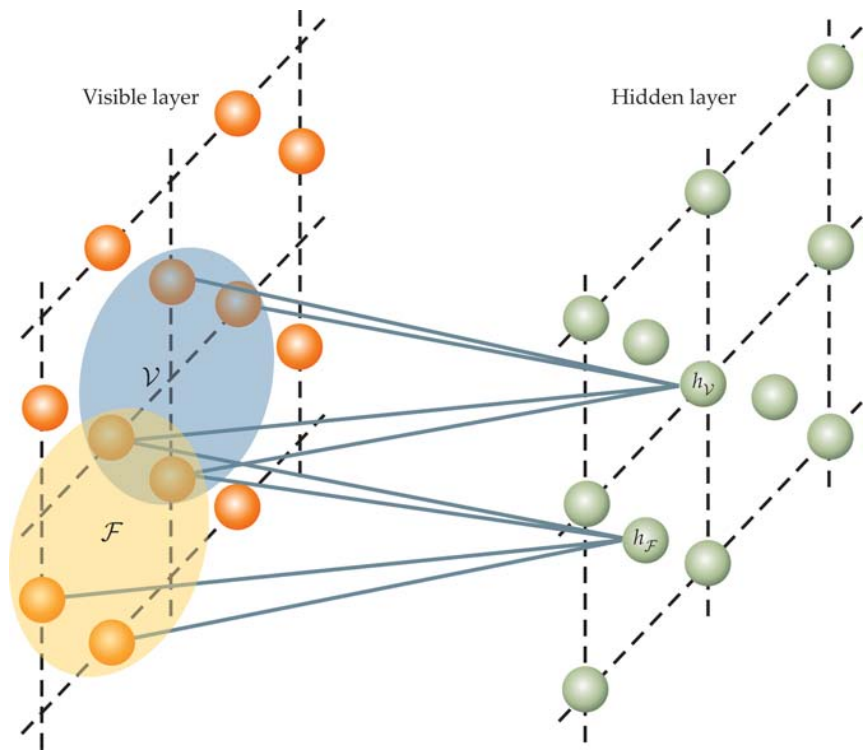

**FIGURE 1. THE RESTRICTED-BOLTZMANN-MACHINE REPRESENTATION** of the toric-code state with intrinsic topological order. Each vertex $\mathcal{V}$ or face $\mathcal{F}$ has four visible neurons that are connected to one hidden neuron $h_\mathcal{V}$ or $h_\mathcal{F}$. The representation is efficient because each connection corresponds to one parameter in the neural network, so the number of parameters scales linearly instead of exponentially with the system size.

# TWO REPRESENTATIONS

A quantum state of a system with $N$ qubits has the general form $|\Psi\rangle = \sum_\Xi \Phi(\Xi)|\Xi\rangle$ where $|\Xi\rangle = (\sigma_1, \sigma_2, \ldots, \sigma_N)$ denotes a possible many-body qubit configuration, and $\Phi(\Xi)$ is a complex function that specifies the amplitude and phase of the state. One can interpret the quantum state as a computational black box that for a given $|\Xi\rangle$ returns a complex number $\Phi(\Xi)$, which is the coefficient for the $|\Xi\rangle$ component of the state.

The tensor-network representation uses tensors to represent quantum states. A tensor's rank indicates its dimensionality, or the number of indices it has, so rank-1 tensors are vectors, rank-2 tensors are matrices, and so on. For simplicity, consider a one-dimensional system with $N$ qubits, shown in panel a, known as the matrix product states (MPS) representation. Each qubit has an associated rank-3 tensor $\mathbf{A}_{ijk}$. The tensors form a network in which the connections represent the indices of the tensors. If two tensors are connected, then their shared index is contracted by summing over all possible values of the repeated index. In the 1D case, two of the indices of each tensor are connected to neighboring tensors and contracted, leaving a rank-1 tensor $\sigma_i$ that represents the physical degrees of freedom. The resulting quantum state is then given by

$$\Phi^{MPS}(\Xi) = \text{Tr}[A_1 A_2 \ldots A_N].$$

A restricted Boltzmann machine (RBM) representation is a neural network with two layers—one visible layer with $N$ visible neurons corresponding to the physical qubits and another layer with $M$ hidden neurons, as shown in panel b. The visible neurons are connected to the hidden neurons, but neurons in the same layer are not connected. The quantum state is given by

$$\Phi^{RBM}(\Xi) = \sum_{\{h\}} e^{\sum a_j \sigma_j + \sum b_k h_k + \sum W_{jk} h_k \sigma_j}$$

where $\{h\}$ denotes the possible configurations $h_1, h_2, \ldots, h_M$ of the hidden neurons, $W_{jk}$ is the coupling strength between the visible and hidden neurons, and $a_j$ and $b_k$ are their bias parameters.

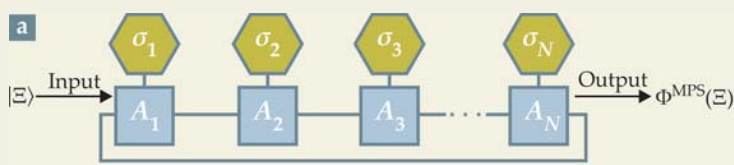
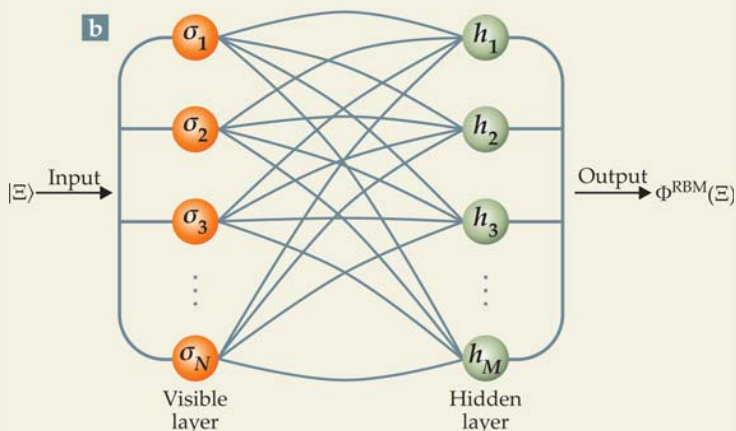

models, the Ising model included, and demonstrated that it could identify various phases and the transitions between them.

## Neural-network representation

In parallel with the fast development of machine-learning algorithms for identifying phases of matter, exciting progress has been made in using artificial neural networks to represent quantum states and solve related quantum many-body problems.

In quantum mechanics, fully describing an arbitrary many-body state requires an exponential amount of information. Consider a system with $N$ qubits, or quantum bits. Each qubit has two possible independent configurations, either 0 or 1; thus there are $2^N$ possible configurations in total. Computationally, that means fully describing the corresponding quantum state requires $2^N$ complex numbers.

The exponential complexity poses an enormous challenge for numerical simulations of quantum many-body systems performed on a classical computer—describing even few qubits requires an extremely large memory. For example, simulating a quantum system with 30 qubits requires tens of gigabytes, about the largest memory for a personal desktop; simulating 50 qubits requires tens of petabytes, more than the memory for the largest supercomputer in the world to date; and simulating 300 qubits requires more bytes than the number of atoms in the observable universe.

Fortunately, most physical states of interest, such as the ground states of many-body Hamiltonians, typically access only a small corner of the entire Hilbert space of quantum states and can therefore be described with a reduced amount of information. Thus designing compact representations of those states in a way that retains their essential physical features is necessary for tackling quantum many-body problems with classical computers.

A renowned description for such states is the tensor-network representation,[7] in which a tensor is assigned to each qubit, and together those tensors describe the many-body quantum state. Such a construction can represent most physical states efficiently in the sense that the amount of information required scales only polynomially, rather than exponentially, with the system size.

Artificial neural networks—highly abstracted and simplified models of the human brain—can also be used to construct compact representations of quantum states. Giuseppe Carleo and Matthias Troyer first explored the idea when they introduced a new representation based on the restricted Boltzmann machine (RBM),[8] a special neural network broadly used in the machine-learning community. (The tensor-network and RBM representations are compared in greater detail in the box above.) An RBM is arranged as two layers of neurons, a visible and a hidden layer, as illustrated in figure 1. The visible neurons represent the physical qubits, and the hidden neurons describe auxiliary degrees of freedom that are eventually eliminated by a summation to produce the network's output, a complex number that serves as the coefficient for the corresponding qubit configuration.

What kinds of quantum many-body states can be efficiently described by RBMs? Certain exotic states, such as topological states, are well represented by RBMs.[9] Figure 1 is a sketch of the RBM representation for the ground state of the toric-code Hamiltonian, a topological state introduced by Alexei Kitaev in the context of topological quantum computation (see the article by one of us [Das Sarma], Michael Freedman, and Chetan Nayak, PHYSICS TODAY, July 2006, page 32). To represent the toric-code state, each hidden neuron of the RBM connects only to its nearest four visible neurons. Each connection is described by one network parameter, so the total number of parameters is roughly four times the number of qubits, which scales linearly, rather than exponentially, with the system size. The strikingly compact representation of the toric-code state can carry over to the excited states as well.

There also exist quantum states with physical interest that carry no efficient RBM description.[10] However, the RBM's applicability increases if it includes an additional hidden layer. The resulting neural network, known as the deep Boltzmann machine, can represent almost all physical quantum states efficiently, with the required number of parameters scaling at most polynomially with the system size.

## Entanglement in neural-network states

What, then, limits neural networks in efficiently representing quantum many-body states? For the conventional tensor-network representation, quantum entanglement is the key. Is it also a critical factor for the neural-network representation?

Quantum entanglement is a physical phenomenon in which measurements on one particle will instantaneously influence the state of another, even when the particles are spatially separated by a large distance—a phenomenon Albert Einstein called "spooky action at a distance." Entanglement is also at the heart of the famous Schrödinger's cat paradox. Both Einstein and Erwin Schrödinger were deeply bothered by quantum entanglement.

Imagine dividing a pure quantum many-body state into two subsystems, A and B, as shown in figure 2. Just as classical many-body systems can be characterized by their entropy, quantum many-body systems can be characterized by their entanglement entropy. Many natural quantum systems satisfy the entanglement area law, which says that the entanglement entropy of a subsystem scales as at most the surface area or the boundary of the subsystem rather than as its volume. That is the case for the Bekenstein–Hawking entropy of a black hole, which scales as the area of its event horizon. In fact, the origin of the black hole entropy is widely believed to be the quantum entanglement between the inside and outside of the black hole. In quantum many-body physics, the ground states of many typical local Hamiltonians also satisfy the entanglement area law, although a rigorous proof of that is notoriously challenging and remains unknown.

The entanglement area law is crucial in the tensor-network representation of quantum many-body states and forms the backbone of numerous tensor-network-based algorithms. In general, the number of parameters that a tensor network needs to describe a quantum state that satisfies the entanglement area law scales only polynomially with the system size. Thus such quantum states typically bear an efficient tensor-network representation. However, for quantum states with massive entanglement, such as highly excited states of quantum Hamiltonians that have volume-law entanglement, the traditional tensor-network representation is not efficient—the number of parameters required scales exponentially with the system size.

All RBM neural-network states with short-range connectivity obey the entanglement area law, independent of their dimensionality and subsystem geometric details.[11] The toric-code states, in which each neuron connects only to its four closest vertices, must then obey the area law, a conclusion that has also been confirmed via other sophisticated mathematical techniques.

Without the short-range condition, general RBM states satisfy an entanglement volume law. In fact, one can analytically construct families of RBM states with maximal entanglement. A sketch of such a construction is shown in figure 2, from which a striking conclusion immediately follows: The RBM description of heavily entangled states is remarkably efficient. Each visible neuron connects to at most three hidden neurons, so the number of parameters scales only linearly with the system size; that scaling demonstrates the unparalleled power of neural networks in describing quantum many-body states with large entanglement. The RBM scaling is in sharp contrast with the traditional tensor-network representation, which requires an exponentially large number of parameters to describe highly entangled states. Clearly, entanglement is not the limiting factor for the efficiency of the neural-network representation.

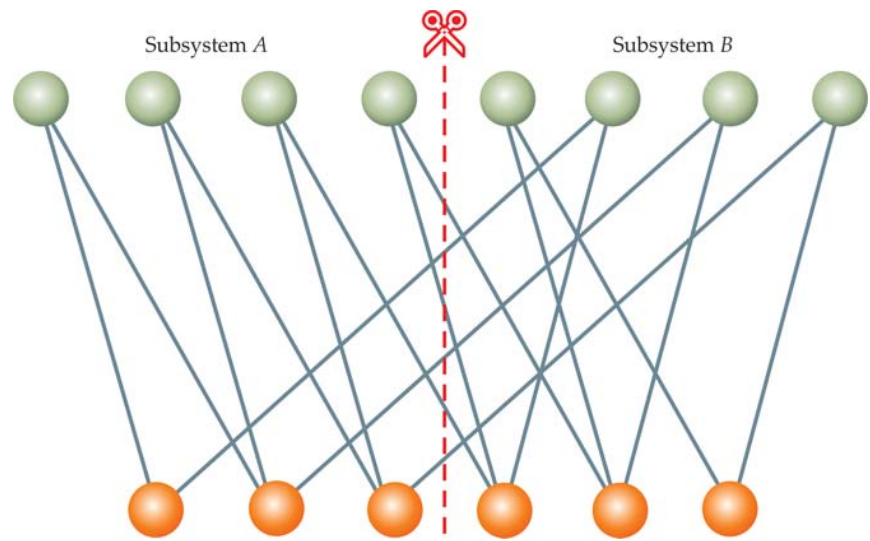

**FIGURE 2. A NEURAL-NETWORK REPRESENTATION** of a one-dimensional quantum state that has maximal volume-law entanglement: If the system is divided into two subsystems, A and B, the entropy of each subsystem is proportional to its volume. Each visible neuron connects to at most three hidden ones, so the number of parameters needed to describe the subsystem scales linearly with the system size rather than exponentially, as in a conventional tensor-network representation.

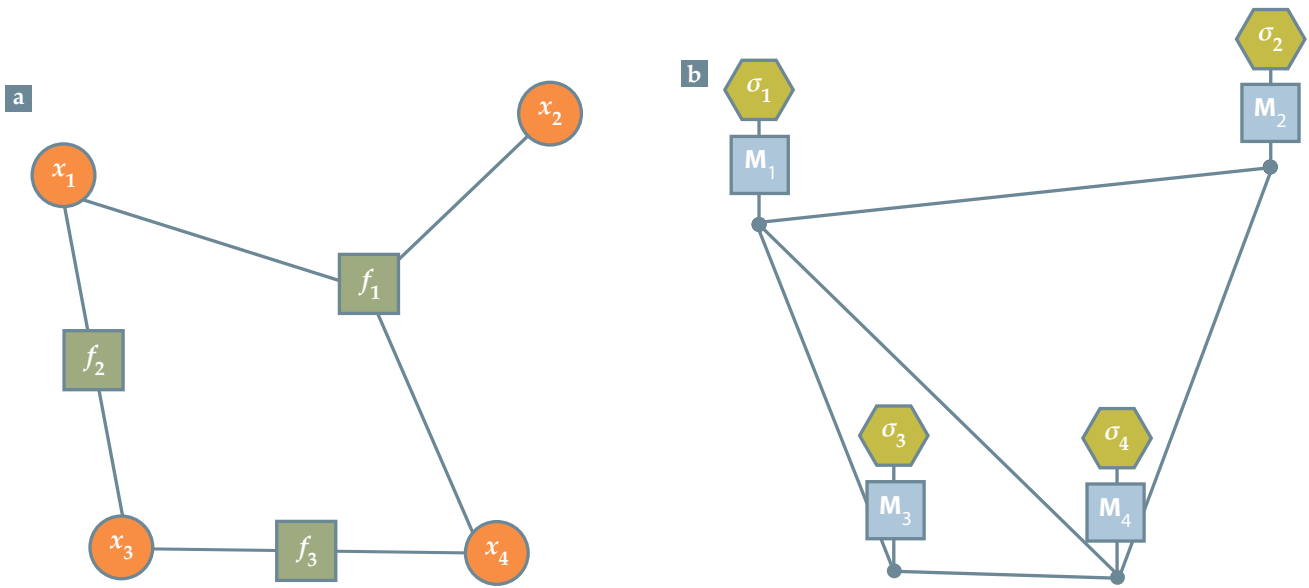

**FIGURE 3. CLASSICAL AND QUANTUM GENERATIVE MODELS** are widely used in both supervised and unsupervised machine learning. **(a)** This illustration of a classical generative model, or factor graph, models the joint probability distribution of the observables $x_i$ as a product of factor functions: $P(x_1, x_2, x_3, x_4) = f_1(x_1, x_2, x_4) f_2(x_1, x_3) f_3(x_3, x_4)$. A classical generative learning task is then reduced to optimizing the adjustable parameters in the factor functions $f_i$. **(b)** This sketch shows a quantum generative model with four qubits $\sigma_i$. The figure represents a special quantum state that is constructed by acting the two-by-two invertible matrices $\mathbf{M}_i$ on a tensor network state. The probability distribution can then be obtained from projective measurements on the resulting state. A quantum generative learning task is then reduced to optimizing the adjustable parameters in the matrices $\mathbf{M}_i$.

## Quantum many-body problems

Solving quantum many-body problems usually entails finding either the system's ground state or the dynamics of the system's time evolution. That can be achieved through an RBM-based variational learning algorithm adopted by Carleo and Troyer in the same paper in which they introduced the RBM representation.[8] They tested the approach on two prototypical quantum spin models—the Ising model in a transverse magnetic field and the antiferromagnetic Heisenberg model—and found that the RBM approach faithfully captured the ground state and the time evolution for each one.

The exceptional ability of neural networks to represent massively entangled states offers a new way to solve intricate many-body problems that involve large entanglement; such problems are challenging or even unsolvable with conventional methods. When applied to a model Hamiltonian with long-range interactions, the RBM-based variational learning algorithm found the system's ground state, which has been numerically shown to hold power-law entanglement.[11] Moreover, the RBM technique has also been used in quantum state tomography—a process of reconstructing the quantum state from the outputs of quantum measurements—for highly entangled states.[12]

Nonlocality, which is closely related to entanglement, is another enigmatic feature of quantum mechanics. As resoundingly established by John Bell, quantum nonlocality precludes any local realistic description of our world and represents the most profound departure of the quantum world from the classical. In practical applications, nonlocality is an indispensable resource for various device-independent quantum technologies, such as secure cryptographic key distribution and certifiable random-number generation. The complete characterization of quantum nonlocality for a generic many-body system is extremely challenging; nevertheless, machine learning, especially RBM-based variational learning, is a promising technique for at least partially solving that problem.[13]

## Quantum-enhanced machine learning

The above examples have clearly uncovered the unparalleled power of machine-learning techniques in solving various challenging quantum problems. Strikingly, unlike in the traditional tensor-network approach, entanglement is not the limiting factor for the efficiency of the neural-network representation and for the related algorithms to learn such a representation. In addition, the neural-network approach works for high-dimensional systems because of the huge flexibility of neural-network structures.

The opposite also holds true: Quantum technologies, especially quantum computing, have the potential to provide a huge boost to machine learning. For one thing, machine learning often deals with large amounts of data, and one common data-analysis technique is the fast Fourier transform (FFT). With quantum computers, there is a quantum version of FFT that is exponentially faster than the classical version.[3] For another, machine-learning algorithms often require solving a huge number of linear problems that amount to doing many matrix multiplications. Quantum computers have intrinsic advantages in executing those operations since quantum mechanics is naturally described by linear algebra—in fact, an early formulation of quantum mechanics by Werner Heisenberg, Max Born, and Pascual Jordan was called matrix mechanics. Thus many conceptual connections exist between machine learning and quantum computing.

Quantum computers aren't expected to speed up every machine-learning algorithm. However, scientists have found a number of quantum algorithms that promise exponential speed increases for certain important tasks.[14] One algorithm that is foundational to the current quantum machine-learning minirevolution is called the HHL algorithm, after its inventors Aram Harrow, Avinatan Hassidim, and Seth Lloyd.[15] Many other quantum learning algorithms either extend HHL or use it as a subroutine. The algorithm seeks to solve a system of linear equations: Given an $N \times N$ matrix **A** and a vector **b**, the aim of HHL is essentially to solve **Ax** = **b** for **x**. For many matrices of physical interest, the HHL algorithm takes on the order of $\log^2 N$ quantum steps to output a quantum state, whereas the best-known classical algorithm requires on the order of $N \log N$ steps.

Several caveats to the HHL algorithm and its variants may nullify its potential benefits.[14] For instance, to map a classical vector to a quantum state the algorithm requires quantum RAM, or qRAM, which could be exponentially expensive. Xun Gao, Zhengyu Zhang, and one of us (Duan) recently introduced a quantum generative model that relaxed the qRAM requirement and thus circumvented the problem of exponential overhead in the initial step of transferring classical data to quantum states.[16]

Compared with more familiar discriminative models, generative models take a different approach to solving problems through machine learning. To understand the difference between the two, consider the earlier example with images of dogs and cats. A discriminative model aims to learn the characteristics that distinguish images of the animals to differentiate between them. The goal of generative models is to be able to produce new images of dogs and cats. In practice, the generative approach is to figure out an underlying probability distribution from a set of training data. In the classical scenario, the probability distribution can be represented by a factor graph. However, for the quantum generative model, the probability distribution is described by a quantum state. Sketches of both the classical and quantum generative models are shown in figure 3.

The quantum generative model has exponential advantages over its classical counterpart in three significant aspects. Not only can it efficiently represent more probability distributions, but the quantum algorithm is also exponentially faster than the classical one both at learning certain probability distributions and at generating new data. The quantum generative model opens a fresh way to explore the power of quantum computing in solving challenging machine-learning problems, and it should thus have important applications in the future.

The above examples are just a glimpse into an increasing zoo of quantum algorithms that may significantly boost machine learning and, more generally, AI tasks.[14] Other intriguing algorithms, such as quantum principal component analysis and quantum support-vector machine, also show great speed-up potentials. In addition, a recently proposed quantum-inspired tensor-network algorithm for machine learning is beginning to show intriguing merits.[17]

## Future partnership

The interdisciplinary field of combining machine learning and quantum physics is growing rapidly, and exciting progress is being made. The above discussions are only the tip of the iceberg.

Applying machine learning to quantum physics requires answers to two crucial questions: What is the killer application for machine learning in solving quantum problems? And can machine learning help discover new physics in quantum systems? An ambitious project that could answer both questions at once is a learning algorithm that specializes in identifying high-$T_c$ superconductors. After training on the huge collection of available experimental data, it should be able to predict new high-$T_c$ superconducting materials and provide new insights into the theory of superconductivity.

For quantum-enhanced machine learning, a unified quantum learning theory has not been developed, and many fundamental questions remain open: What is the general criterion for determining if a machine-learning task can be significantly expedited by a quantum computer? What learning problems can be efficiently solved by a quantum computer but not by a classical one? And how can a quantum computer efficiently analyze large quantum data sets that may eventually be available?

For classical machine learning, there is an exact map between the variational renormalization-group method in physics—an iterative coarse-graining scheme that extracts relevant features for a physical system at different length scales—and deep learning,[18] and that map gives valuable insight on why deep learning is powerful. Is it possible to construct such a map for the case of quantum deep learning? Moreover, a smoking-gun experimental demonstration of quantum speed-ups in a practical machine-learning task would be an important milestone.

It is hard to foresee when the first practical quantum computer will be available and harder still to predict what the quantum future will look like. Yet one thing is certain: The marriage of machine learning and quantum physics is a symbiotic relationship that could transform them both.